\documentclass[aps,prb,twocolumn,groupedaddress,showpacs]{revtex4-1}

\usepackage{graphicx}

\begin{document}

\title{Magnetic Single Domain State of the Monochiral Helimagnet MnSi \\in Zero Field Limit: Magnetic Properties Study}

\author{V. N. Narozhnyi}
\email{narozhnyivn@gmail.com}
\affiliation{Institute for High Pressure Physics, Russian Academy of Sciences, 142190, Troitsk, Moscow, Russia}
\author{V. N. Krasnorussky}
\altaffiliation{Present address: A. M. Prokhorov General Physics Institute of the RAS, 38 Vavilov Street, 119991 Moscow, Russia.}
\affiliation{Institute for High Pressure Physics, Russian Academy of Sciences, 142190, Troitsk, Moscow, Russia}


\date{\today}

\begin{abstract}
Attention is drawn to a possibility to obtain the monochiral helimagnet MnSi in a magnetic single domain state (SDS) \emph{in zero magnetic field limit}. It is shown that this metastable zero field magnetic SDS can be achieved by gradual decrease of the field down to zero after initial transformation of MnSi to a spin-polarized state in high fields $H$. This can be achieved only for $\bf{H}\parallel$ [111]. Investigations of MnSi in magnetic SDS give us a possibility to determine the two components of low field magnetic susceptibility of this compound, $\chi_{\perp}(T)$ and $\chi_{\parallel}(T)$ [$\chi_{\perp}$ and $\chi_{\parallel}$ correspond to $\bf{H}\perp$\, and $\bf{H}\parallel$ (111)-plane containing magnetic moments in helically ordered state]. These results are compared with macroscopic magnetic susceptibility of MnSi earlier determined for this compound only in magnetic multi domain state (MDS). In addition our results are compared with the data reported for some non-monochiral helimagnets. Characteristic features of \emph{monochiral} helimagnets are elucidated.
\end{abstract}


\maketitle

\section{Introduction}
Chirality is one of the basic properties of a nature. If a substance is chiral, its  various properties may be influenced by chirality. Magnetic properties of chiral magnetic structures (helimagnets) were extensively investigated since 1959 \cite{Yoshimori_1959}. At the beginning \textit{non-monochiral} helimagnets  were studied, e.g., metal oxides of $\text{MnO}_2$ type and heavy rare earth metals (having equal population of magnetic domains with left and right associated handedness). Then investigations of a \textit{monochiral} helimagnets (e.g., MnSi) with quite different origin of chirality \cite{Nakanishi_1980} began. In this Paper we try to elucidate (on an example of MnSi) some characteristic features inherent to magnetic behavior of \textit{monochiral} helimagnets.

MnSi is a long-period helimagnet with chiral magnetic structure formed below transition temperature $T_{\text{N}}\approx29\,\text{K}$ \cite{Ishikawa_1976,Motoya_1976}. Magnetic structure of MnSi is actually \textit{monochiral} due to its rigorous relation with particular crystallographic handedness of each MnSi sample \cite{PhysRevB.28.6251,Ishida_1985,PhysRevB.84.014435}. Magnetic ordering \cite{Williams_1966,Wernick_1972,Levinson_1973,Kusaka_1976,Hansen_1977,PhysRevB.55.8330,PhysRevB.74.214414,PhysRevB.73.174402,PhysRevB.82.064404,PhysRevB.85.214418} as well as magnetic phase diagram \cite{Kusaka_1976,Ishikawa_1984,Lebech_1995,PhysRevB.74.214414,PhysRevB.73.174402,PhysRevB.82.064404,PhysRevB.85.214418,Our_Conf_1,Our_Conf} of MnSi and some related B20 noncentrosymmetric cubic compounds were extensively studied. Magnetic helix in MnSi is characterized by wave vector $\bf{Q}$ directed in magnetic field $H=0$ along [111], magnetic moments being located in ferromagnetic (FM) planes perpendicular to $\bf{Q}$. Directions of magnetic moments in adjacent FM (111)-type planes are turned by small angle forming a long-period helix.  A model based on an hierarchy of three types of interactions [ferromagnetic isotropic exchange interaction; antisymmetric Dzyaloshinskii-Moriya interaction (DMI), connected with lack of inversion symmetry; and anisotropic exchange interaction] describes main features of magnetic ordering of these compounds fairly well \cite{Nakanishi_1980,Bak_1980}.

Previous magnetization studies have shown that $M(H)$ of MnSi in the ordered state is close to linear for $H\leq H_{C\text{2}}(T)$ and abruptly changes slope in higher fields \cite{Williams_1966,Wernick_1972,Levinson_1973,Bloch_1975,Hansen_1977,PhysRevB.62.986,PhysRevB.82.064404,PhysRevB.85.214418}. This sharp turn is a consequence of a transformation to a spin-polarized ("induced ferromagnetic") state at $H>H_{C\text{2}}$. Some anomalies were observed in $M(H)$ dependencies of MnSi at much smaller fields $H_{C\text{1}}(T)$ which are usually attributed to formation of magnetic \textit{single domain state} (SDS) by rotating of $\bf{Q}$ of all magnetic domains towards $\bf{H}$ \cite{Ishikawa_1976,Plumer_1981,Kataoka_1981,PhysRevB.74.214414,PhysRevB.73.174402} (four directions of $\bf{Q}$ of the type [111] are crystallographically equivalent). A decrease of $H$ (after transformation to SDS) leads usually to (partial) reconstruction of magnetic \textit{multidomain state} (MDS) \cite{Ishikawa_1976}. To the best of our knowledge no indication on a possibility to obtain MnSi in magnetic SDS in the zero field limit was reported so far. Magnetization study of MnSi described in this Paper unexpectedly shows that: (1) magnetic SDS of MnSi can be obtained at $H=0$; (2) this can be achieved only for $\bf{H}$ initially parallel to [111].

At temperatures few degrees above $T_{\text{N}}$ (namely $T\gtrsim33\,\text{K}$) MnSi behaves like a typical paramagnet (PM) above FM Curie temperature. Magnetic susceptibility $\chi(T)$ of MnSi is well described by Curie-Weiss law $\chi(T) = \text{C}/(T - \theta)$ with \textit{positive} $\theta\approx T_{\text{N}}$ and effective magnetic moment $\mu_{\text{eff}}\simeq2.2\,\mu_{\text{B}}$ (saturation magnetic moment is appreciably smaller, $\simeq0.4\,\mu_{\text{B}}$) \cite{Wernick_1972,Levinson_1973}. Only just above $T_{\text{N}}$ $\chi(T)$ deviates considerably from this dependence (diverging at $T=\theta$) and remains finite at $T=T_{\text{N}}$. In the ordered state of MnSi $\chi(T)$ decreases rapidly just below $T_{\text{N}}$ and flattens at lower $T$ forming an anomaly clearly visible around $T_{\text{N}}$. This rather sharp anomaly in $\chi(T)$ is basically considered as an intrinsic for MnSi, see, e.g., \cite{Kusaka_1976,PhysRevB.55.8330,PhysRevB.82.064404,PhysRevB.87.134407}.

In this Paper it is shown that $\chi(T)$ dependence described above is actually connected with a magnetic MDS of MnSi samples for which an averaged macroscopic magnetic susceptibility was usually measured, see, e.g., \cite{Kusaka_1976,PhysRevB.55.8330,PhysRevB.82.064404,PhysRevB.87.134407}. Determination of the two components of magnetic susceptibility $\chi_{\perp}(T)$ and $\chi_{\parallel}(T)$ (corresponding to $\bf{H}\perp$ and $\bf{H}\parallel$ (111)-plane containing magnetic moments \footnote{Our use of $\chi_{\perp}$ and $\chi_{\parallel}$ corresponds to generally accepted for other types of magnetic materials. For MnSi $\chi_{\parallel}$ corresponds to $\bf{H}$ parallel to (111)-plane, containing magnetic moments, i.e., $\bf{H} \perp \bf{Q}$. In some papers, see, e.g., Ref.\,\onlinecite{PhysRevB.73.174402}, $\chi_{\parallel}$ corresponds to $\bf{H} \parallel \bf{Q}$}) of MnSi in SDS gives results quite different from the described above for the case of MnSi in MDS. $\chi_{\perp}(T)$ remains practically constant below $T_{\text{N}}$ whereas $\chi_{\parallel}(T)$ experiences a clear decrease. This behavior of magnetic susceptibility of MnSi resembls to some extent a behavior of two components of magnetic susceptibility of collinear easy axis antiferromagnets (AFM). This result is significant not only for understanding intrinsic magnetic properties of MnSi as well as a behavior of a single magnetic helix in this compound but also can be an important background for a description of the so called A-phase having a complicated structure of helixes \cite{Kusaka_1976,PhysRevB.73.174402,Rosler_2006}.

The obtained results are also compared with ones reported for some other magnetic materials including \emph{non-monochiral} helimagnets whose magnetic ordering is not generated by DMI.

\section{Experimental}
A cubic sample with faces parallel to (110), (1$\bar{1}$0) and (001) (edges $\approx3$\,mm) was cut from MnSi single crystal batch grown by Bridgman technique. Magnetization was measured by "Lake Shore Cryotronics" vibrating sample magnetometer with transverse configuration of magnetic field and a possibility to rotate a sample holder. $M(H)$ measurements were performed at $5.5\,\text{K}\leq T \leq 35\,\text{K}$ for $\bf{H} \parallel$ [111], [001] and [110]. $M(H)$ was usually measured after zero field cooling with $H$ varying from 0 to 11\,kOe and back to 0. Temperature was measured with accuracy better then 0.1\,K. Temperature stability during $M(H)$ runs was better then 0.05\,K. Magnetic ordering temperature $T_{\text{N}}$ (determined at a maximum in $\partial \chi(T)/\partial T$ slope, see MDS data in Fig.~\ref{Fig4}) was found to be 28.8\,K, which coincides with $T_{\text{N}}$ determined in Ref.\,\onlinecite{Stishov_2008} at a maximum of specific heat of another sample cut from the same batch.

\section{Results and Discussion}
\begin{figure} [tb]
\includegraphics{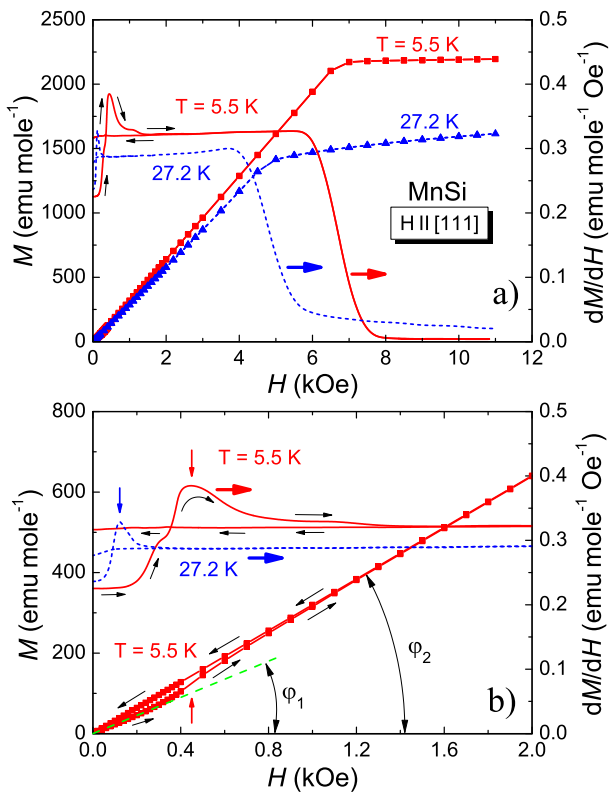}%
\caption{\label{Fig1} (Color online) (a) Magnetization $M(H)$ (left axis) and $\partial M/\partial H(H)$ (right axis, pointed by thick arrows for clarity) of MnSi for $\bf{H}\parallel\text{[111]}$ at $T=\text{5.5~K}$ - full line (red online) and at $T=27.2~\text{K}$  - dashed line (blue online). Full squares (red online) and
full triangles (blue online) represent the experimental data, the lines connecting symbols are guides for eye. The sequence of $H$ variation is shown by thin arrows for $\partial M/\partial H(H)$ and $T=5.5~\text{K}$ only. (b) Expanded view for $H\leq2~\text{kOe}$ [$M(H)$ is shown for $T=5.5~\text{K}$ only]. Vertical arrows indicate fields $H_{C\text{1}}$. The dashed line at panel b) (green online) represents initial $M(H)$ slope for increasing field.}
\end{figure}

\begin{figure} [tb]
\includegraphics{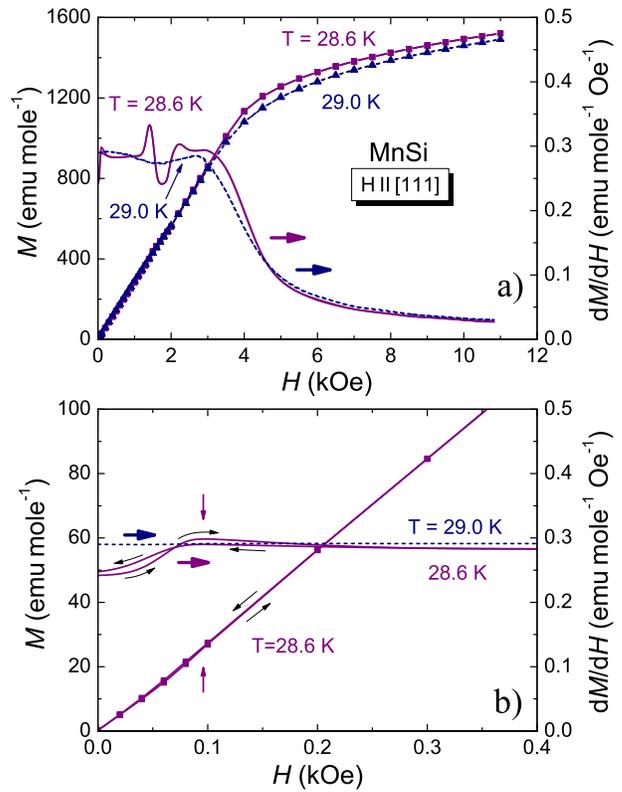}%
\caption{\label{Fig2} (Color online) (a) Magnetization $M(H)$ (left axis) and $\partial M/\partial H(H)$ (right axis) of MnSi for $\bf{H} \parallel \text{[111]}$ at $T=\text{28.6~K}=T_{\text{N}}-\text{0.2~K}$ - full line (purple online) and at $T=\text{29.0~K}=T_{\text{N}}+\text{0.2~K}$ - dashed line (navy online). (b) Expanded view for $H\leq0.4~\text{kOe}$ [$M(H)$ is shown for $T=28.6~\text{K}$ only]. Details are the same as for the Fig.~\ref{Fig1}.}
\end{figure}

\begin{figure} [tb]
\includegraphics{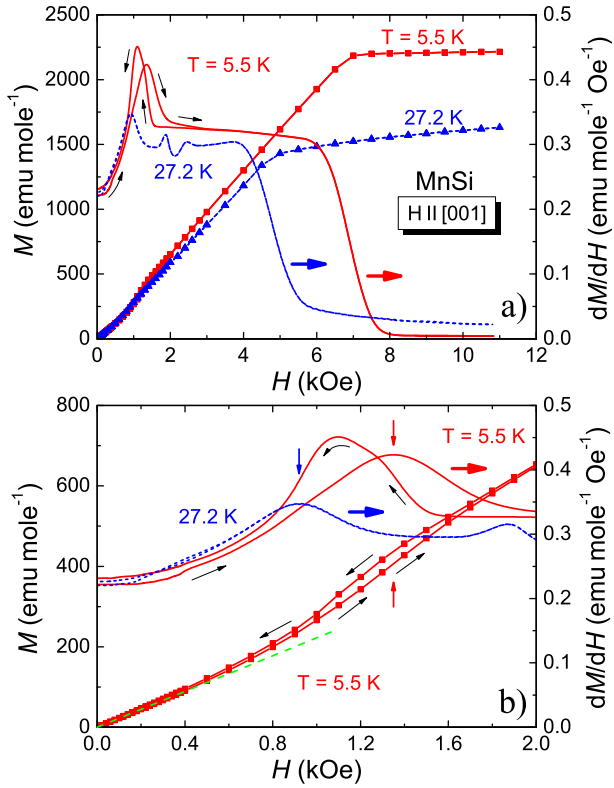}%
\caption{\label{Fig3} (Color online) (a) Magnetization $M(H)$ (left axis) and $\partial M/\partial H(H)$ (right axis) of MnSi for $\bf{H} \parallel \text{[001]}$ at $T=\text{5.5~K}$ - full line (red online) and at $T=\text{27.2~K}$ - dashed line (blue online). (b) Expanded view for $H\leq2~\text{kOe}$ [$M(H)$ is shown for $T=5.5~\text{K}$ only]. Details are the same as for Fig.~\ref{Fig1}.}
\end{figure}

Some representative examples of $M(H)$ curves for $\bf{H} \parallel$ [111] are shown in Figs.~\ref{Fig1} and \ref{Fig2}. $M(H)$ dependencies determined for $\bf{H} \parallel$ [001] at the same temperatures as in Fig.~\ref{Fig1} are shown in Fig.~\ref{Fig3}. Upper panels of figures show $M(H)$ as well as $\partial M(H)/\partial H$ up to the maximal field. [$\partial M(H)/\partial H$ was determined by numerical differentiating of $M(H)$. The level of uncertainty of this derivative is rather small due to a smooth variation of magnetic moment with field and high precision of magnetization measurements. It may exceed the width of corresponding lines in these figures only in several times. Magnetic field steps used in the experiments give a possibility to catch clearly all the observed anomalies in the $M(H)$ and $\partial M(H)/\partial H$ dependencies.] Lower panels show expanded views for smaller fields. Results for $\bf{H} \parallel$ [110] are qualitatively similar to ones for $\bf{H} \parallel$ [001] and have not been shown for brevity.

$M(H)$ dependencies are close to linear at $H\leq(4 - 6)\,\text{kOe}$ and abruptly change slopes for higher fields. This is connected with transformation to the spin-polarized state \cite{Williams_1966,Wernick_1972,Levinson_1973,Bloch_1975,Hansen_1977,PhysRevB.62.986,PhysRevB.82.064404,PhysRevB.85.214418}. Some nonlinearities can be also seen in $M(H)$ for smaller $H$ (see lower panels of Figs.~\ref{Fig1}\,-\,\ref{Fig3}). More clearly these nonlinearities are visible on $\partial M(H)/\partial H$ curves. Maxima in $\partial M(H)/\partial H$ (denoted in lower panels by vertical arrows) is related with transformation of MnSi to magnetic SDS (earlier this transformation was observed by neutron diffraction \cite{PhysRevB.74.214414}). Although anomalies in comparable fields were observed by various methods \cite{Matsunaga_1982,Thessieu_1997,PhysRevB.62.986,PhysRevB.73.224440,PhysRevB.74.214414}, our systematic study of $M(H)$ in wide temperature range and for all three principal directions of magnetic field has shown that $H_{C\text{1}}$ depends essentially on temperature as well as on the direction of magnetic field (it varies from, e.g., $80~\text{Oe}$ for $T=27.5~\text{K}$ and $\bf{H} \parallel$ [111] to $1.3~\text{kOe}$ for $T=5.5~\text{K}$ and $\bf{H} \parallel$ [001]). On the basis of these results magnetic phase diagrams of MnSi were substantially refined especially at low $H$, see Refs.\,\onlinecite{Our_Conf_1,Our_Conf} for details.

For the purposes of this paper it is important that $M(H)$ and $\partial M(H)/\partial H$ dependencies are essentially \textit{irreversible} at $5.5\leq T \leq 27.4\,\text{K}$ for $\bf{H} \parallel$ [111], see Fig.~\ref{Fig1}. For $T = 5.5\,\text{K}$ virgin $M(H)$ curve, starting from $H=0$, has a smaller slope than $M(H)$ for decreasing $H$ (after reaching $H=11\,\text{kOe}$), the value of $\partial M(H)/\partial H$ being practically constant for decreasing field for $H < 6~\text{kOe}$. The ratio of the two slopes $\tan\varphi_1/\tan\varphi_2$ for $T=5.5~\text{K}$ and $\bf{H} \parallel$ [111] [(1) - for virgin $M(H)$ curve and (2) - for decreasing $H$] equals $0.71 \pm 0.02$. Constant slope of $M(H)$ for decreasing $H$ means obviously that the sample remains in magnetic SDS down to $H=0$.

Magnetic behavior remains qualitatively very similar for $5.5\leq T \leq 27.4\,\text{K}$, Fig.~\ref{Fig1}, but changes drastically in the vicinity of $T_{\text{N}}$, see Fig.~\ref{Fig2}. For $27.6\leq T \leq 28.8\,\text{K}=T_{\text{N}}$ $M(H)$ curves became practically \textit{reversible}. Only a slight irreversibility in low fields can be seen at $T = 28.6\,\text{K}$  for $\partial M(H)/\partial H$ curve shown in Fig.~\ref{Fig2}b. That means that close to $T_{\text{N}}$ the magnetic SDS experiences back transformation to MDS with decrease of magnetic field even for $\bf{H} \parallel$ [111]. As expected, magnetization is perfectly reversible at temperatures above $T_{\text{N}}$, even at $T=29.0\,\text{K}=T_{\text{N}}+0.2\,\text{K}$, see Fig.~\ref{Fig2}b.

For temperatures just below $T_{\text{N}}$ (e.g., for $T=28.6\,\text{K}$, Fig.~\ref{Fig2}a) two other anomalies appear for $\partial M(H)/\partial H$ dependence at $H \approx 1.2$ and 2.0\,kOe. These anomalies are connected with crossing the border of so-called A-phase, see Refs.\,\onlinecite{Kusaka_1976,Our_Conf}. Interestingly that for $\bf{H} \parallel$ [111] the temperature interval for which reversible behavior was observed in the ordered state at small fields practically coincides with temperatures for which A-phase can be seen an higher fields. Probably this coincidence is not accidental and may be connected with intrinsic behavior of magnetic helices in this compound. Magnetic behavior of MnSi in the A-phase will be described and compared with k-flop \cite{PhysRevB.73.174402} as well as with skyrmion \cite{Rosler_2006} models elsewhere.

At the same time magnetic behavior of MnSi is close to \textit{reversible} at all temperatures for two other principal directions of magnetic field, $\bf{H} \parallel$ [001] (Fig.~\ref{Fig3}) and $\bf{H} \parallel$ [110] (not shown in the figures), especially in low fields. This is quite different from the results for $\bf{H} \parallel$ [111] described above. Almost reversible $M(H)$ for these directions of magnetic field means that, after initial transformation to magnetic SDS, decrease of $H$ leads to the back transformation to MDS. The fact that MnSi can be obtained in magnetic SDS in the zero field limit for $\bf{H} \parallel$ [111] only is connected most probably with [111]-type orientation of the helix wave vector $\bf{Q}$ at $H=0$ in this compound. At the same time an increase of magnetic fluctuations at temperatures just below $T_{\text{N}}$ can lead to \textit{reversible} behavior at these temperatures  even for $\bf{H} \parallel$ [111] due to a global instability of SDS in MnSi near $T_{\text{N}}$.

Although magnetic behavior of MnSi below $T_{\text{N}}$ depends to some extent on the direction of $\bf{H}$, $M(H)$ curves at low fields are qualitatively similar to ones proposed in Ref.\,\onlinecite{Plumer_1981} (and Ref.\,\onlinecite{Kataoka_1981} for the case of low anisotropy) for helicoid with $\bf{Q}$ gradually rotating with increase of field to the direction of $\bf{H}$. Comparison of our results with these models as well as with the data reported earlier for the related compounds Fe(Co)Si will be discussed elsewhere.

\begin{figure} [tb]
\includegraphics{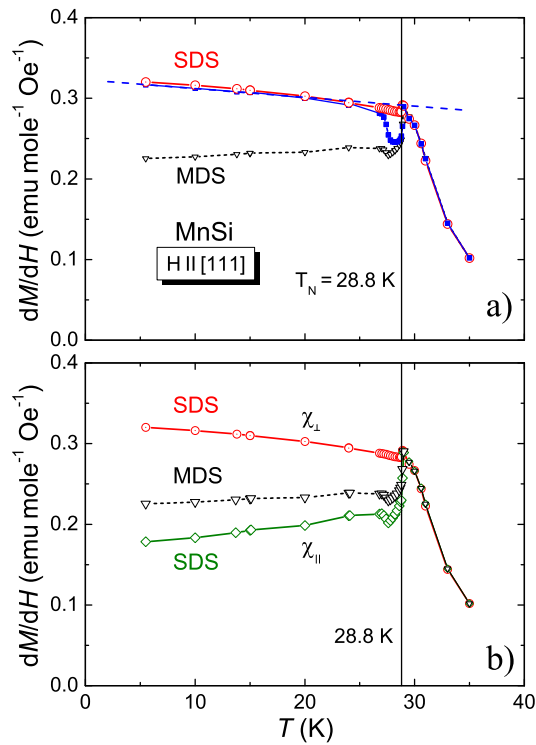}%
\caption{\label{Fig4} (Color online) Magnetic properties of MnSi for $\bf{H} \parallel \text{[111]}$. (a) $\partial M/\partial H(T)$ determined at $H \rightarrow 0$ for \textit{increasing} field - virgin curve (open black triangles connected by dotted line); $\partial M/\partial H(T)$ determined at $H \rightarrow 0$ for \textit{decreasing} field - back run after $H=11~\text{kOe}$ (full squares connected by solid line, blue online); $\partial M/\partial H(T)$ determined at $H=400~\text{Oe}$ ($H=400~\text{Oe} \gg H_{C\text{1}} \simeq 80~\text{Oe}$ for $27.5~\text{K} \lesssim T < T_{\text{N}}$) for \textit{decreasing} field (open circles connected by solid line, red online). Dashed straight line (blue online) is drawn through full squares at $5.5\leq T \leq 20~\text{K}$ and extrapolated to higher $T$. (b) $\partial M/\partial H(T)$ of MnSi in a SDS for $\bf{H} \perp \text{(111)}$-plane [open circles connected by solid line (red online) - the same data as in the panel a)] and $\bf{H} \parallel \text{(111)}$-plane [open diamonds connected by solid line, olive online]. Results for MDS is also shown by open black triangles connected by dotted line.}
\end{figure}

Obtaining of MnSi in magnetic SDS gives a possibility to determine a $\chi_{\perp}(T)$ (corresponding to $\bf{H}$ perpendicular to the magnetic planes, i.e. $\bf{H} \parallel \bf{Q}$, see Ref.\,28). For this purpose it is necessary to determine correctly the values of $\partial M(H)/\partial H$ for MnSi in SDS, e.g., by using $H\rightarrow 0$ for \textit{decreasing} fields at $\bf{H} \parallel$ [111] after initial transformation to the spin polarized state in high fields. The results are shown in Fig.~\ref{Fig4}a by full squares (blue online).

The dip in this dependence just below $T_{\text{N}}$ is worthy of notice. At the same time linear extrapolation of $\chi_{\perp}(T)$ obtained from the data at $T \leq 20\,\text{K}$ (shown by dashed line) intersects with $\chi_{\perp}(T)$ in paramagnetic state at $T \approx 29.0\,\text{K}$, i.e., very close to $T_{\text{N}}$. This gives an idea that this dip may be connected with an incorrect determination of $\chi_{\perp}(T)$ in the narrow region just below $T_{\text{N}}$, namely at $27.6\leq T \leq 28.8\,\text{K}$ [related simply with \textit{reversibility} in $M(H)$ of MnSi at these temperatures and, therefore, MDS instead of the required for a correct determination of $\chi_{\perp}(T)$ single domain state], correct $\chi_{\perp}(T)$ being close to the linear extrapolation described above.

To demonstrate this it is necessary to determine correctly $\chi_{\perp}(T)$ not only at $5.5\leq T \leq 27.6\,\text{K}$ [as described above for region with irreversible $M(H)$], but also at $27.6\leq T \leq 28.8\,\text{K}$ with reversible $M(H)$ behavior. It can be done for this temperature interval  from $M(H)$ data at $H>H_{C\text{1}}(T)\simeq80\,\text{Oe}$, i.e., at fields sufficiently high for MnSi to be in SDS. We have used $H=400\,\text{Oe}$ for this procedure [$\chi_{\perp}(T)=\partial M(H)/\partial H$, shown by open circles (red online) on Fig.\,\ref{Fig4}a] because transition to magnetic SDS is rather broad [see $\partial M(H)/\partial H$ curve for $T=27.2\,\text{K}$ on Fig.\,\ref{Fig1}b].

Values of $\chi_{\perp}(T)$ determined by the two methods described above are very close for $T \leq 24\,\text{K}$ and practically coincide in paramagnetic state, see Fig.~\ref{Fig4}a. Therefore the data shown by open circles (red online) in Fig.~\ref{Fig4}a can be considered as $\chi_{\perp}(T)$ dependence really intrinsic for MnSi.

The next step is a determination of $\chi_{\parallel}(T)$ (corresponding to $\bf{H}\parallel$ (111)-plane, i.e., $\bf{H} \perp \bf{Q}$, see Ref.\,28). $\chi_{\parallel}(T)$ can be easily determined from $\chi_{\perp}(T)$ and the averaged macroscopic magnetic susceptibility $\chi(T)$ (obtained in magnetic MDS) using equation $\chi(T)=1/3\chi_{\perp}(T)+2/3\chi_{\parallel}(T)$ valid for virgin magnetic MDS of cubic single crystal (for any direction of $\bf{H}$) as well as for polycrystalline MnSi. Results are shown in Fig.\,\ref{Fig4}b by open diamonds (olive online).

Thereby both components of magnetic susceptibility of MnSi, $\chi_{\perp}(T)$ and $\chi_{\parallel}(T)$, in magnetic SDS have been experimentally determined for $5.5\leq T \leq T_{\text{N}}$. Pay attention to a considerable difference with $\chi(T)$ obtained in magnetic MDS (shown by open down triangles) that was considered as intrinsic for MnSi for a long time (since the very first measurements of $\chi(T)$ of MnSi, see, e.g., Refs. \onlinecite{Kusaka_1976,PhysRevB.55.8330,PhysRevB.82.064404}).

It is interesting to note that the ratio $\chi_{\parallel}/\chi_{\perp}\simeq0.56$ at $T=5.5\,\text{K}$ corresponds to the ratio of $M(H)$ slopes in MDS and SDS $\tan\varphi_1/\tan\varphi_2=0.71$ (see Figs.~\ref{Fig1} and \ref{Fig3}) observed for $\bf{H} \parallel$ [111] as well as for $\bf{H} \parallel$ [001]. This is connected with the existence of four types equally populated magnetic domains with $\bf{Q}$ vectors directed along four space diagonals of MnSi cube structure. For $\bf{H} \parallel$ [111] magnetization of the domain with $\bf{Q}\parallel$ [111] is determined by $\chi_{\perp}$ only, whereas magnetization of each of three other types of domains is determined by both components of $\chi$. Elemental calculation in the assumption of equally populated magnetic domains gives the value of the ratio of $M(H)$ slopes 0.71 if $\chi_{\parallel}/\chi_{\perp} = 0.56$.

\begin{figure} [tb]
\includegraphics{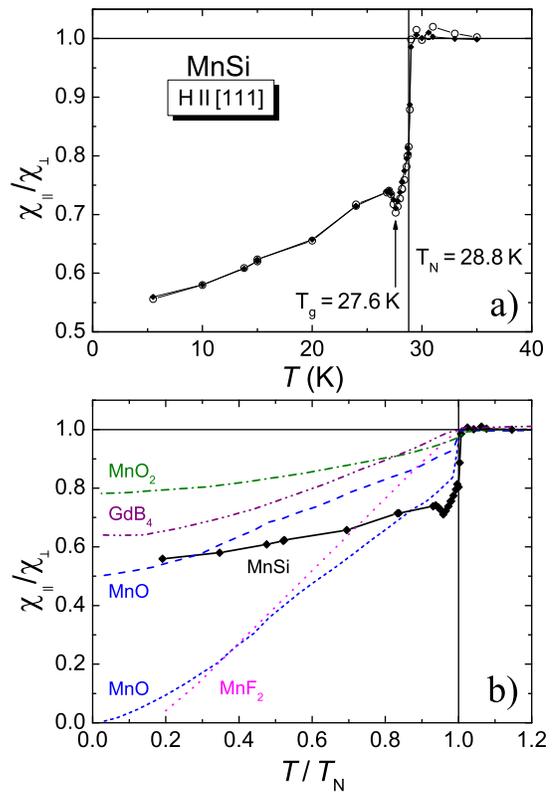}%
\caption{\label{Fig5} (Color online) (a) $\chi_{\parallel}/\chi_{\perp}$ \textit{vs}. $T$ for MnSi. $\chi_{\perp}(T)$ was determined as (1) $\partial M(H)/\partial H$ at $H=400\,\text{Oe}$ (open circles) and (2) extrapolated to $H=0$ $\partial M(H)/\partial H$ data in SDS (full circles). $M(H)$ back runs data at $\bf{H} \parallel$ [111] were used in both cases. (b) Comparison of  $\chi_{\parallel}/\chi_{\perp}$ \textit{vs} $T/T_{\text{N}}$ for MnSi with data reported for $\text{MnF}_{2}$ \cite{Bizette_1954}, MnO  (upper curve - directly measured results, lower - derived for collinear structure taking into account existence of $S$ domains \cite {PhysRevB.23.1185}), $\text{GdB}_{4}$ \cite{Cho_2005}, $\text{MnO}_{2}$ \cite{PhysRevB.61.3563}. Data for MnSi are the same as in the Fig.~\ref{Fig5}a.}
\end{figure}

The ratio $\chi_{\parallel}/\chi_{\perp}$ \textit{vs}. $T$ is shown in Fig.\,\ref{Fig5}a. This ratio decreases dramatically near $T_{\text{N}}$ and reaches a value of 0.7 at $T=27.6\,\text{K}=T_{\text{g}}$. A nature of the small dip in $\chi_{\parallel}(T)/\chi_{\perp}(T)$ at $T=T_{\text{g}}$ is not clear so far. It could be mentioned only that a reversible behavior of $M(H)$ for $\bf{H} \parallel$ [111] in small fields as well as an appearance of A-phase in higher fields are observed at the same temperature interval $(T_{\text{g}} - T_{\text{N}})$.

The drop in $\chi_{\parallel}/\chi_{\perp}$ \textit{vs}. $T$ is connected with steep decrease of $\chi_{\parallel}(T)$ and practically constant $\chi_{\perp}(T)$ near $T_{\text{N}}$. At further cooling (below $T_{\text{g}}$) $\chi_{\parallel}(T)/\chi_{\perp}(T)$ decreases more gradually and reaches 0.56 at $T=5.5\,\text{K}$. Sharp decrease of $\chi_{\parallel}(T)$ below $T_{\text{N}}$ can be characterized as a rapid increase of perpendicular ($\bf{H} \perp \bf{Q}$) magnetic stiffness of a helix. Temperature interval $(T_{\text{g}} - T_{\text{N}})$ may be considered as a region where magnetic helix is forming when $T$ decreases near $T_{\text{N}}$. As far as we know there is no theory describing a $\chi(T)$ dependence for \textit{monochiral} helimagnets. In Ref.\,\onlinecite{PhysRevB.73.174402} it is only pointed out that perpendicular (with respect to $\bf{Q}$) component of $\chi$ [$\chi_{\parallel}(T)$ in our notation] should be 0.5 of the value of longitudinal one for $T=0$ and that this is connected with DMI interaction. Our results for the ratio $\chi_{\parallel}/\chi_{\perp}\simeq0.56$ at $T=5.5\,\text{K}$ are rather close to one obtained in this theory \cite{PhysRevB.73.174402}.

It is of interest to compare the dependence of $\chi_{\parallel}/\chi_{\perp}$ \textit{vs}. $T$ observed for MnSi with the data reported for some materials with different types of magnetic order [see Fig.\,\ref{Fig5}b] of which $\text{MnF}_{2}$ \cite{Bizette_1954} is a collinear easy axis AFM; MnO - collinear planar AFM for which existence of $S$ (spin rotation) domains should be taken into account \cite {PhysRevB.23.1185}; $\text{GdB}_{4}$ - non-collinear planar AFM \cite{Cho_2005}, and $\text{MnO}_{2}$ - helicoidal (\emph{non-monochiral}) magnet whose ordering is described by a model with three exchange integrals \cite{PhysRevB.61.3563}. The behavior of MnSi near and just below $T_{\text{N}}$ is very different from ones of these compounds, see Fig.\,\ref{Fig5}b. At $T/T_{\text{N}} \rightarrow 0$ $\chi_{\parallel}/\chi_{\perp} \rightarrow 0$ for systems with a collinear magnetic ordering. For non-collinear magnetic structures $\chi_{\parallel}/\chi_{\perp}$ ratios may be even comparable to 1.

A model describing $\chi_{\parallel}/\chi_{\perp}$ \textit{vs}. $T$ for antiferromagnets from general grounds has been reported recently \cite{PhysRevLett.109.077201,arXiv:1407.6353}. It describes the observed dependencies of $\chi_{\parallel}/\chi_{\perp}$ \textit{vs}. $T$ for collinear and non-collinear AFMs quite well. For non-collinear orderings at $T=0$ this model gives $\chi_{\parallel}/\chi_{\perp}$ from 0.1 to 1. For helicoidal $\text{MnO}_{2}$ it gives $\approx 0.76$ that is close to the observed 0.78 (see Ref.\,\onlinecite{PhysRevB.61.3563}). An attempt to apply formally the results of Ref.\,\onlinecite{PhysRevLett.109.077201} to MnSi gives $\chi_{\parallel}/\chi_{\perp}$ close to 0.1 due to the long period of magnetic helicoid in this compound. This differs a lot from the observed value of 0.56. It is one more (in addition to the behavior of $\chi_{\parallel}/\chi_{\perp}$ near $T_{\text{N}}$ as well as to a tendency to form an A-phase) indication on a different nature of \textit{monochiral} helical order in MnSi in comparison with, e.g., \textit{non-monochiral} $\text{MnO}_{2}$.

\section{Conclusion}
In summary, it was demonstrated that a magnetic single domain state in MnSi can be obtained even \emph{in the zero magnetic field limit}. Measurements in SDS allowed us to determine (for the first time) two principal components of magnetic susceptibility  $\chi_{\parallel}(T)$ and $\chi_{\perp}(T)$ of MnSi. The obtained results are quite different from the behavior of an averaged macroscopic magnetic susceptibility of this compound reported so far. The observed dependence of $\chi_{\parallel}/\chi_{\perp}$ \textit{vs}. $T$ is proved to be very different from the data reported for magnets with dissimilar types of magnetic ordering (including non-monochiral helical order). This difference elucidate the characteristic feature of MnSi probably inherent also for other  \emph{monochiral} helimagnets.

\begin{acknowledgments}
We are grateful to A. I. Smirnov for enlightening discussions. We thank S. M. Stishov for a possibility to use MnSi samples for measurements.
\end{acknowledgments}

\bibliography{MnSi_single_domain}

\providecommand{\noopsort}[1]{}\providecommand{\singleletter}[1]{#1}%
\begin{thebibliography}{39}%
\makeatletter
\providecommand \@ifxundefined [1]{%
 \@ifx{#1\undefined}
}%
\providecommand \@ifnum [1]{%
 \ifnum #1\expandafter \@firstoftwo
 \else \expandafter \@secondoftwo
 \fi
}%
\providecommand \@ifx [1]{%
 \ifx #1\expandafter \@firstoftwo
 \else \expandafter \@secondoftwo
 \fi
}%
\providecommand \natexlab [1]{#1}%
\providecommand \enquote  [1]{``#1''}%
\providecommand \bibnamefont  [1]{#1}%
\providecommand \bibfnamefont [1]{#1}%
\providecommand \citenamefont [1]{#1}%
\providecommand \href@noop [0]{\@secondoftwo}%
\providecommand \href [0]{\begingroup \@sanitize@url \@href}%
\providecommand \@href[1]{\@@startlink{#1}\@@href}%
\providecommand \@@href[1]{\endgroup#1\@@endlink}%
\providecommand \@sanitize@url [0]{\catcode `\\12\catcode `\$12\catcode
  `\&12\catcode `\#12\catcode `\^12\catcode `\_12\catcode `\%12\relax}%
\providecommand \@@startlink[1]{}%
\providecommand \@@endlink[0]{}%
\providecommand \url  [0]{\begingroup\@sanitize@url \@url }%
\providecommand \@url [1]{\endgroup\@href {#1}{\urlprefix }}%
\providecommand \urlprefix  [0]{URL }%
\providecommand \Eprint [0]{\href }%
\providecommand \doibase [0]{http://dx.doi.org/}%
\providecommand \selectlanguage [0]{\@gobble}%
\providecommand \bibinfo  [0]{\@secondoftwo}%
\providecommand \bibfield  [0]{\@secondoftwo}%
\providecommand \translation [1]{[#1]}%
\providecommand \BibitemOpen [0]{}%
\providecommand \bibitemStop [0]{}%
\providecommand \bibitemNoStop [0]{.\EOS\space}%
\providecommand \EOS [0]{\spacefactor3000\relax}%
\providecommand \BibitemShut  [1]{\csname bibitem#1\endcsname}%
\let\auto@bib@innerbib\@empty
\bibitem [{\citenamefont {Yoshimori}(1959)}]{Yoshimori_1959}%
  \BibitemOpen
  \bibfield  {author} {\bibinfo {author} {\bibfnamefont {A.}~\bibnamefont
  {Yoshimori}},\ }\href@noop {} {\bibfield  {journal} {\bibinfo  {journal} {J.
  Phys. Soc. Japan}\ }\textbf {\bibinfo {volume} {14}},\ \bibinfo {pages} {807}
  (\bibinfo {year} {1959})}\BibitemShut {NoStop}%
\bibitem [{\citenamefont {Nakanishi}\ \emph {et~al.}(1980)\citenamefont
  {Nakanishi}, \citenamefont {Yanase}, \citenamefont {Hasegawa},\ and\
  \citenamefont {Kataoka}}]{Nakanishi_1980}%
  \BibitemOpen
  \bibfield  {author} {\bibinfo {author} {\bibfnamefont {O.}~\bibnamefont
  {Nakanishi}}, \bibinfo {author} {\bibfnamefont {A.}~\bibnamefont {Yanase}},
  \bibinfo {author} {\bibfnamefont {A.}~\bibnamefont {Hasegawa}}, \ and\
  \bibinfo {author} {\bibfnamefont {M.}~\bibnamefont {Kataoka}},\ }\href@noop
  {} {\bibfield  {journal} {\bibinfo  {journal} {Solid. State. Commun.}\
  }\textbf {\bibinfo {volume} {35}},\ \bibinfo {pages} {995} (\bibinfo {year}
  {1980})}\BibitemShut {NoStop}%
\bibitem [{\citenamefont {Ishikawa}\ \emph {et~al.}(1976)\citenamefont
  {Ishikawa}, \citenamefont {Tajima}, \citenamefont {Bloch},\ and\
  \citenamefont {Roth}}]{Ishikawa_1976}%
  \BibitemOpen
  \bibfield  {author} {\bibinfo {author} {\bibfnamefont {Y.}~\bibnamefont
  {Ishikawa}}, \bibinfo {author} {\bibfnamefont {K.}~\bibnamefont {Tajima}},
  \bibinfo {author} {\bibfnamefont {D.}~\bibnamefont {Bloch}}, \ and\ \bibinfo
  {author} {\bibfnamefont {M.}~\bibnamefont {Roth}},\ }\href@noop {} {\bibfield
   {journal} {\bibinfo  {journal} {Solid. State. Commun.}\ }\textbf {\bibinfo
  {volume} {19}},\ \bibinfo {pages} {525} (\bibinfo {year} {1976})}\BibitemShut
  {NoStop}%
\bibitem [{\citenamefont {Motoya}\ \emph {et~al.}(1976)\citenamefont {Motoya},
  \citenamefont {Yasuoka}, \citenamefont {Nakamura},\ and\ \citenamefont
  {Wernick}}]{Motoya_1976}%
  \BibitemOpen
  \bibfield  {author} {\bibinfo {author} {\bibfnamefont {K.}~\bibnamefont
  {Motoya}}, \bibinfo {author} {\bibfnamefont {H.}~\bibnamefont {Yasuoka}},
  \bibinfo {author} {\bibfnamefont {Y.}~\bibnamefont {Nakamura}}, \ and\
  \bibinfo {author} {\bibfnamefont {J.}~\bibnamefont {Wernick}},\ }\href@noop
  {} {\bibfield  {journal} {\bibinfo  {journal} {Solid. State. Commun.}\
  }\textbf {\bibinfo {volume} {19}},\ \bibinfo {pages} {529} (\bibinfo {year}
  {1976})}\BibitemShut {NoStop}%
\bibitem [{\citenamefont {Shirane}\ \emph {et~al.}(1983)\citenamefont
  {Shirane}, \citenamefont {Cowley}, \citenamefont {Majkrzak}, \citenamefont
  {Sokoloff}, \citenamefont {Pagonis}, \citenamefont {Perry},\ and\
  \citenamefont {Ishikawa}}]{PhysRevB.28.6251}%
  \BibitemOpen
  \bibfield  {author} {\bibinfo {author} {\bibfnamefont {G.}~\bibnamefont
  {Shirane}}, \bibinfo {author} {\bibfnamefont {R.}~\bibnamefont {Cowley}},
  \bibinfo {author} {\bibfnamefont {C.}~\bibnamefont {Majkrzak}}, \bibinfo
  {author} {\bibfnamefont {J.~B.}\ \bibnamefont {Sokoloff}}, \bibinfo {author}
  {\bibfnamefont {B.}~\bibnamefont {Pagonis}}, \bibinfo {author} {\bibfnamefont
  {C.~H.}\ \bibnamefont {Perry}}, \ and\ \bibinfo {author} {\bibfnamefont
  {Y.}~\bibnamefont {Ishikawa}},\ }\href {\doibase 10.1103/PhysRevB.28.6251}
  {\bibfield  {journal} {\bibinfo  {journal} {Phys. Rev. B}\ }\textbf {\bibinfo
  {volume} {28}},\ \bibinfo {pages} {6251} (\bibinfo {year}
  {1983})}\BibitemShut {NoStop}%
\bibitem [{\citenamefont {Ishida}\ \emph {et~al.}(1985)\citenamefont {Ishida},
  \citenamefont {Endoh}, \citenamefont {Mitsuda}, \citenamefont {Ishikawa},\
  and\ \citenamefont {Tanaka}}]{Ishida_1985}%
  \BibitemOpen
  \bibfield  {author} {\bibinfo {author} {\bibfnamefont {M.}~\bibnamefont
  {Ishida}}, \bibinfo {author} {\bibfnamefont {Y.}~\bibnamefont {Endoh}},
  \bibinfo {author} {\bibfnamefont {S.}~\bibnamefont {Mitsuda}}, \bibinfo
  {author} {\bibfnamefont {Y.}~\bibnamefont {Ishikawa}}, \ and\ \bibinfo
  {author} {\bibfnamefont {M.}~\bibnamefont {Tanaka}},\ }\href@noop {}
  {\bibfield  {journal} {\bibinfo  {journal} {J. Phys. Soc. Japan}\ }\textbf
  {\bibinfo {volume} {54}},\ \bibinfo {pages} {2975} (\bibinfo {year}
  {1985})}\BibitemShut {NoStop}%
\bibitem [{\citenamefont {Dyadkin}\ \emph {et~al.}(2011)\citenamefont
  {Dyadkin}, \citenamefont {Grigoriev}, \citenamefont {Menzel}, \citenamefont
  {Chernyshov}, \citenamefont {Dmitriev}, \citenamefont {Schoenes},
  \citenamefont {Maleyev}, \citenamefont {Moskvin},\ and\ \citenamefont
  {Eckerlebe}}]{PhysRevB.84.014435}%
  \BibitemOpen
  \bibfield  {author} {\bibinfo {author} {\bibfnamefont {V.~A.}\ \bibnamefont
  {Dyadkin}}, \bibinfo {author} {\bibfnamefont {S.~V.}\ \bibnamefont
  {Grigoriev}}, \bibinfo {author} {\bibfnamefont {D.}~\bibnamefont {Menzel}},
  \bibinfo {author} {\bibfnamefont {D.}~\bibnamefont {Chernyshov}}, \bibinfo
  {author} {\bibfnamefont {V.}~\bibnamefont {Dmitriev}}, \bibinfo {author}
  {\bibfnamefont {J.}~\bibnamefont {Schoenes}}, \bibinfo {author}
  {\bibfnamefont {S.~V.}\ \bibnamefont {Maleyev}}, \bibinfo {author}
  {\bibfnamefont {E.~V.}\ \bibnamefont {Moskvin}}, \ and\ \bibinfo {author}
  {\bibfnamefont {H.}~\bibnamefont {Eckerlebe}},\ }\href {\doibase
  10.1103/PhysRevB.84.014435} {\bibfield  {journal} {\bibinfo  {journal} {Phys.
  Rev. B}\ }\textbf {\bibinfo {volume} {84}},\ \bibinfo {pages} {014435}
  (\bibinfo {year} {2011})}\BibitemShut {NoStop}%
\bibitem [{\citenamefont {Williams}\ \emph {et~al.}(1966)\citenamefont
  {Williams}, \citenamefont {Wernick}, \citenamefont {Sherwood},\ and\
  \citenamefont {Wertheim}}]{Williams_1966}%
  \BibitemOpen
  \bibfield  {author} {\bibinfo {author} {\bibfnamefont {H.~J.}\ \bibnamefont
  {Williams}}, \bibinfo {author} {\bibfnamefont {J.~H.}\ \bibnamefont
  {Wernick}}, \bibinfo {author} {\bibfnamefont {R.~C.}\ \bibnamefont
  {Sherwood}}, \ and\ \bibinfo {author} {\bibfnamefont {G.~K.}\ \bibnamefont
  {Wertheim}},\ }\href@noop {} {\bibfield  {journal} {\bibinfo  {journal} {J.
  Appl. Phys.}\ }\textbf {\bibinfo {volume} {37}},\ \bibinfo {pages} {1256}
  (\bibinfo {year} {1966})}\BibitemShut {NoStop}%
\bibitem [{\citenamefont {Wernick}\ \emph {et~al.}(1972)\citenamefont
  {Wernick}, \citenamefont {Wertheim},\ and\ \citenamefont
  {Sherwood}}]{Wernick_1972}%
  \BibitemOpen
  \bibfield  {author} {\bibinfo {author} {\bibfnamefont {J.~H.}\ \bibnamefont
  {Wernick}}, \bibinfo {author} {\bibfnamefont {G.~K.}\ \bibnamefont
  {Wertheim}}, \ and\ \bibinfo {author} {\bibfnamefont {R.~C.}\ \bibnamefont
  {Sherwood}},\ }\href@noop {} {\bibfield  {journal} {\bibinfo  {journal} {Mat.
  Res. Bull.}\ }\textbf {\bibinfo {volume} {7}},\ \bibinfo {pages} {1431}
  (\bibinfo {year} {1972})}\BibitemShut {NoStop}%
\bibitem [{\citenamefont {Levinson}\ \emph {et~al.}(1973)\citenamefont
  {Levinson}, \citenamefont {Lander},\ and\ \citenamefont
  {Steinitz}}]{Levinson_1973}%
  \BibitemOpen
  \bibfield  {author} {\bibinfo {author} {\bibfnamefont {L.~M.}\ \bibnamefont
  {Levinson}}, \bibinfo {author} {\bibfnamefont {G.~H.}\ \bibnamefont
  {Lander}}, \ and\ \bibinfo {author} {\bibfnamefont {M.~O.}\ \bibnamefont
  {Steinitz}},\ }\href@noop {} {\bibfield  {journal} {\bibinfo  {journal} {AIP
  Conf. Proc.}\ }\textbf {\bibinfo {volume} {10}},\ \bibinfo {pages} {1138}
  (\bibinfo {year} {1973})}\BibitemShut {NoStop}%
\bibitem [{\citenamefont {Kusaka}\ \emph {et~al.}(1976)\citenamefont {Kusaka},
  \citenamefont {Yamamoto}, \citenamefont {Komatsubara},\ and\ \citenamefont
  {Ishikawa}}]{Kusaka_1976}%
  \BibitemOpen
  \bibfield  {author} {\bibinfo {author} {\bibfnamefont {S.}~\bibnamefont
  {Kusaka}}, \bibinfo {author} {\bibfnamefont {K.}~\bibnamefont {Yamamoto}},
  \bibinfo {author} {\bibfnamefont {T.}~\bibnamefont {Komatsubara}}, \ and\
  \bibinfo {author} {\bibfnamefont {Y.}~\bibnamefont {Ishikawa}},\ }\href@noop
  {} {\bibfield  {journal} {\bibinfo  {journal} {Solid. State. Commun.}\
  }\textbf {\bibinfo {volume} {20}},\ \bibinfo {pages} {925} (\bibinfo {year}
  {1976})}\BibitemShut {NoStop}%
\bibitem [{\citenamefont {Hansen}(1977)}]{Hansen_1977}%
  \BibitemOpen
  \bibfield  {author} {\bibinfo {author} {\bibfnamefont {P.}~\bibnamefont
  {Hansen}},\ }\href@noop {} {}\bibinfo {type} {{Ris{\o} Report No.}}\ \bibinfo
  {number} {360}\ (\bibinfo {year} {1977})\BibitemShut {NoStop}%
\bibitem [{\citenamefont {Pfleiderer}\ \emph {et~al.}(1997)\citenamefont
  {Pfleiderer}, \citenamefont {McMullan}, \citenamefont {Julian},\ and\
  \citenamefont {Lonzarich}}]{PhysRevB.55.8330}%
  \BibitemOpen
  \bibfield  {author} {\bibinfo {author} {\bibfnamefont {C.}~\bibnamefont
  {Pfleiderer}}, \bibinfo {author} {\bibfnamefont {G.~J.}\ \bibnamefont
  {McMullan}}, \bibinfo {author} {\bibfnamefont {S.~R.}\ \bibnamefont
  {Julian}}, \ and\ \bibinfo {author} {\bibfnamefont {G.~G.}\ \bibnamefont
  {Lonzarich}},\ }\href {\doibase 10.1103/PhysRevB.55.8330} {\bibfield
  {journal} {\bibinfo  {journal} {Phys. Rev. B}\ }\textbf {\bibinfo {volume}
  {55}},\ \bibinfo {pages} {8330} (\bibinfo {year} {1997})}\BibitemShut
  {NoStop}%
\bibitem [{\citenamefont {Grigoriev}\ \emph
  {et~al.}(2006{\natexlab{a}})\citenamefont {Grigoriev}, \citenamefont
  {Maleyev}, \citenamefont {Okorokov}, \citenamefont {Chetverikov},
  \citenamefont {B\"oni}, \citenamefont {Georgii}, \citenamefont {Lamago},
  \citenamefont {Eckerlebe},\ and\ \citenamefont
  {Pranzas}}]{PhysRevB.74.214414}%
  \BibitemOpen
  \bibfield  {author} {\bibinfo {author} {\bibfnamefont {S.~V.}\ \bibnamefont
  {Grigoriev}}, \bibinfo {author} {\bibfnamefont {S.~V.}\ \bibnamefont
  {Maleyev}}, \bibinfo {author} {\bibfnamefont {A.~I.}\ \bibnamefont
  {Okorokov}}, \bibinfo {author} {\bibfnamefont {Y.~O.}\ \bibnamefont
  {Chetverikov}}, \bibinfo {author} {\bibfnamefont {P.}~\bibnamefont {B\"oni}},
  \bibinfo {author} {\bibfnamefont {R.}~\bibnamefont {Georgii}}, \bibinfo
  {author} {\bibfnamefont {D.}~\bibnamefont {Lamago}}, \bibinfo {author}
  {\bibfnamefont {H.}~\bibnamefont {Eckerlebe}}, \ and\ \bibinfo {author}
  {\bibfnamefont {K.}~\bibnamefont {Pranzas}},\ }\href {\doibase
  10.1103/PhysRevB.74.214414} {\bibfield  {journal} {\bibinfo  {journal} {Phys.
  Rev. B}\ }\textbf {\bibinfo {volume} {74}},\ \bibinfo {pages} {214414}
  (\bibinfo {year} {2006}{\natexlab{a}})}\BibitemShut {NoStop}%
\bibitem [{\citenamefont {Maleyev}(2006)}]{PhysRevB.73.174402}%
  \BibitemOpen
  \bibfield  {author} {\bibinfo {author} {\bibfnamefont {S.~V.}\ \bibnamefont
  {Maleyev}},\ }\href {\doibase 10.1103/PhysRevB.73.174402} {\bibfield
  {journal} {\bibinfo  {journal} {Phys. Rev. B}\ }\textbf {\bibinfo {volume}
  {73}},\ \bibinfo {pages} {174402} (\bibinfo {year} {2006})}\BibitemShut
  {NoStop}%
\bibitem [{\citenamefont {Bauer}\ \emph {et~al.}(2010)\citenamefont {Bauer},
  \citenamefont {Neubauer}, \citenamefont {Franz}, \citenamefont {M\"unzer},
  \citenamefont {Garst},\ and\ \citenamefont
  {Pfleiderer}}]{PhysRevB.82.064404}%
  \BibitemOpen
  \bibfield  {author} {\bibinfo {author} {\bibfnamefont {A.}~\bibnamefont
  {Bauer}}, \bibinfo {author} {\bibfnamefont {A.}~\bibnamefont {Neubauer}},
  \bibinfo {author} {\bibfnamefont {C.}~\bibnamefont {Franz}}, \bibinfo
  {author} {\bibfnamefont {W.}~\bibnamefont {M\"unzer}}, \bibinfo {author}
  {\bibfnamefont {M.}~\bibnamefont {Garst}}, \ and\ \bibinfo {author}
  {\bibfnamefont {C.}~\bibnamefont {Pfleiderer}},\ }\href {\doibase
  10.1103/PhysRevB.82.064404} {\bibfield  {journal} {\bibinfo  {journal} {Phys.
  Rev. B}\ }\textbf {\bibinfo {volume} {82}},\ \bibinfo {pages} {064404}
  (\bibinfo {year} {2010})}\BibitemShut {NoStop}%
\bibitem [{\citenamefont {Bauer}\ and\ \citenamefont
  {Pfleiderer}(2012)}]{PhysRevB.85.214418}%
  \BibitemOpen
  \bibfield  {author} {\bibinfo {author} {\bibfnamefont {A.}~\bibnamefont
  {Bauer}}\ and\ \bibinfo {author} {\bibfnamefont {C.}~\bibnamefont
  {Pfleiderer}},\ }\href {\doibase 10.1103/PhysRevB.85.214418} {\bibfield
  {journal} {\bibinfo  {journal} {Phys. Rev. B}\ }\textbf {\bibinfo {volume}
  {85}},\ \bibinfo {pages} {214418} (\bibinfo {year} {2012})}\BibitemShut
  {NoStop}%
\bibitem [{\citenamefont {Kadowaki}\ \emph {et~al.}(1982)\citenamefont
  {Kadowaki}, \citenamefont {Okuda},\ and\ \citenamefont
  {Date}}]{Ishikawa_1984}%
  \BibitemOpen
  \bibfield  {author} {\bibinfo {author} {\bibfnamefont {K.}~\bibnamefont
  {Kadowaki}}, \bibinfo {author} {\bibfnamefont {K.}~\bibnamefont {Okuda}}, \
  and\ \bibinfo {author} {\bibfnamefont {M.}~\bibnamefont {Date}},\ }\href@noop
  {} {\bibfield  {journal} {\bibinfo  {journal} {J. Phys. Soc. Japan}\ }\textbf
  {\bibinfo {volume} {51}},\ \bibinfo {pages} {2433} (\bibinfo {year}
  {1982})}\BibitemShut {NoStop}%
\bibitem [{\citenamefont {Lebech}\ \emph {et~al.}(1995)\citenamefont {Lebech},
  \citenamefont {Harrisa}, \citenamefont {Pedersena}, \citenamefont
  {Mortensena}, \citenamefont {Gregoryb}, \citenamefont {Bernhoeftc},
  \citenamefont {Jermyd},\ and\ \citenamefont {Browne}}]{Lebech_1995}%
  \BibitemOpen
  \bibfield  {author} {\bibinfo {author} {\bibfnamefont {B.}~\bibnamefont
  {Lebech}}, \bibinfo {author} {\bibfnamefont {P.}~\bibnamefont {Harrisa}},
  \bibinfo {author} {\bibfnamefont {J.~S.}\ \bibnamefont {Pedersena}}, \bibinfo
  {author} {\bibfnamefont {K.}~\bibnamefont {Mortensena}}, \bibinfo {author}
  {\bibfnamefont {C.}~\bibnamefont {Gregoryb}}, \bibinfo {author}
  {\bibfnamefont {N.}~\bibnamefont {Bernhoeftc}}, \bibinfo {author}
  {\bibfnamefont {M.}~\bibnamefont {Jermyd}}, \ and\ \bibinfo {author}
  {\bibfnamefont {S.}~\bibnamefont {Browne}},\ }\href@noop {} {\bibfield
  {journal} {\bibinfo  {journal} {J. Magn. Magn. Mater.}\ }\textbf {\bibinfo
  {volume} {140-144}},\ \bibinfo {pages} {119} (\bibinfo {year}
  {1995})}\BibitemShut {NoStop}%
\bibitem [{\citenamefont {Narozhnyi}\ and\ \citenamefont
  {Krasnorussky}(2012)}]{Our_Conf_1}%
  \BibitemOpen
  \bibfield  {author} {\bibinfo {author} {\bibfnamefont {V.~N.}\ \bibnamefont
  {Narozhnyi}}\ and\ \bibinfo {author} {\bibfnamefont {V.~N.}\ \bibnamefont
  {Krasnorussky}},\ }in\ \href@noop {} {\emph {\bibinfo {booktitle} {Extended
  Abstracts of the 36th Meeting on Low Temperature Physics}}},\ \bibinfo
  {address} {St.-Petersburg, Russia}\ (\bibinfo {organization} {Ioffe PTI
  RAS},\ \bibinfo {year} {2012})\ pp.\ \bibinfo {pages} {65--66},\ \bibinfo
  {note} {(in Russian, unpublished)},\ \Eprint
  {http://arxiv.org/abs/\text{http://www.ioffe.ru/nt36/main\_menu/nt36ba.pdf}}
  {\text{http://www.ioffe.ru/nt36/main\_menu/nt36ba.pdf}} \BibitemShut
  {NoStop}%
\bibitem [{\citenamefont {Narozhnyi}\ and\ \citenamefont
  {Krasnorussky}(2013)}]{Our_Conf}%
  \BibitemOpen
  \bibfield  {author} {\bibinfo {author} {\bibfnamefont {V.~N.}\ \bibnamefont
  {Narozhnyi}}\ and\ \bibinfo {author} {\bibfnamefont {V.~N.}\ \bibnamefont
  {Krasnorussky}},\ }\href@noop {} {\bibfield  {journal} {\bibinfo  {journal}
  {Zh. Eksp. Teor. Fiz.}\ }\textbf {\bibinfo {volume} {143}},\ \bibinfo {pages}
  {906} (\bibinfo {year} {2013})},\ \translation{Sov. Phys. JETP \textbf{116},
  785 (2013)}\BibitemShut {NoStop}%
\bibitem [{\citenamefont {Bak}\ and\ \citenamefont {Jensen}(1980)}]{Bak_1980}%
  \BibitemOpen
  \bibfield  {author} {\bibinfo {author} {\bibfnamefont {P.}~\bibnamefont
  {Bak}}\ and\ \bibinfo {author} {\bibfnamefont {M.~H.}\ \bibnamefont
  {Jensen}},\ }\href@noop {} {\bibfield  {journal} {\bibinfo  {journal} {J.
  Phys. C}\ }\textbf {\bibinfo {volume} {13}},\ \bibinfo {pages} {L881}
  (\bibinfo {year} {1980})}\BibitemShut {NoStop}%
\bibitem [{\citenamefont {Bloch}\ \emph {et~al.}(1975)\citenamefont {Bloch},
  \citenamefont {Voiron}, \citenamefont {Jaccarino},\ and\ \citenamefont
  {Wernick}}]{Bloch_1975}%
  \BibitemOpen
  \bibfield  {author} {\bibinfo {author} {\bibfnamefont {D.}~\bibnamefont
  {Bloch}}, \bibinfo {author} {\bibfnamefont {J.}~\bibnamefont {Voiron}},
  \bibinfo {author} {\bibfnamefont {V.}~\bibnamefont {Jaccarino}}, \ and\
  \bibinfo {author} {\bibfnamefont {J.~H.}\ \bibnamefont {Wernick}},\
  }\href@noop {} {\bibfield  {journal} {\bibinfo  {journal} {Phys. Lett.}\
  }\textbf {\bibinfo {volume} {51A}},\ \bibinfo {pages} {259} (\bibinfo {year}
  {1975})}\BibitemShut {NoStop}%
\bibitem [{\citenamefont {Koyama}\ \emph {et~al.}(2000)\citenamefont {Koyama},
  \citenamefont {Goto}, \citenamefont {Kanomata},\ and\ \citenamefont
  {Note}}]{PhysRevB.62.986}%
  \BibitemOpen
  \bibfield  {author} {\bibinfo {author} {\bibfnamefont {K.}~\bibnamefont
  {Koyama}}, \bibinfo {author} {\bibfnamefont {T.}~\bibnamefont {Goto}},
  \bibinfo {author} {\bibfnamefont {T.}~\bibnamefont {Kanomata}}, \ and\
  \bibinfo {author} {\bibfnamefont {R.}~\bibnamefont {Note}},\ }\href {\doibase
  10.1103/PhysRevB.62.986} {\bibfield  {journal} {\bibinfo  {journal} {Phys.
  Rev. B}\ }\textbf {\bibinfo {volume} {62}},\ \bibinfo {pages} {986} (\bibinfo
  {year} {2000})}\BibitemShut {NoStop}%
\bibitem [{\citenamefont {Plumer}\ and\ \citenamefont
  {Walker}(1981)}]{Plumer_1981}%
  \BibitemOpen
  \bibfield  {author} {\bibinfo {author} {\bibfnamefont {M.~L.}\ \bibnamefont
  {Plumer}}\ and\ \bibinfo {author} {\bibfnamefont {M.~B.}\ \bibnamefont
  {Walker}},\ }\href@noop {} {\bibfield  {journal} {\bibinfo  {journal} {J.
  Phys. C: Solid State Phys.}\ }\textbf {\bibinfo {volume} {14}},\ \bibinfo
  {pages} {4689} (\bibinfo {year} {1981})}\BibitemShut {NoStop}%
\bibitem [{\citenamefont {Kataoka}\ and\ \citenamefont
  {Nakanishi}(1981)}]{Kataoka_1981}%
  \BibitemOpen
  \bibfield  {author} {\bibinfo {author} {\bibfnamefont {M.}~\bibnamefont
  {Kataoka}}\ and\ \bibinfo {author} {\bibfnamefont {O.}~\bibnamefont
  {Nakanishi}},\ }\href@noop {} {\bibfield  {journal} {\bibinfo  {journal} {J.
  Phys. Soc. Japan}\ }\textbf {\bibinfo {volume} {50}},\ \bibinfo {pages}
  {3888} (\bibinfo {year} {1981})}\BibitemShut {NoStop}%
\bibitem [{\citenamefont {Janoschek}\ \emph {et~al.}(2013)\citenamefont
  {Janoschek}, \citenamefont {Garst}, \citenamefont {Bauer}, \citenamefont
  {Krautscheid}, \citenamefont {Georgii}, \citenamefont {B\"oni},\ and\
  \citenamefont {Pfleiderer}}]{PhysRevB.87.134407}%
  \BibitemOpen
  \bibfield  {author} {\bibinfo {author} {\bibfnamefont {M.}~\bibnamefont
  {Janoschek}}, \bibinfo {author} {\bibfnamefont {M.}~\bibnamefont {Garst}},
  \bibinfo {author} {\bibfnamefont {A.}~\bibnamefont {Bauer}}, \bibinfo
  {author} {\bibfnamefont {P.}~\bibnamefont {Krautscheid}}, \bibinfo {author}
  {\bibfnamefont {R.}~\bibnamefont {Georgii}}, \bibinfo {author} {\bibfnamefont
  {P.}~\bibnamefont {B\"oni}}, \ and\ \bibinfo {author} {\bibfnamefont
  {C.}~\bibnamefont {Pfleiderer}},\ }\href {\doibase
  10.1103/PhysRevB.87.134407} {\bibfield  {journal} {\bibinfo  {journal} {Phys.
  Rev. B}\ }\textbf {\bibinfo {volume} {87}},\ \bibinfo {pages} {134407}
  (\bibinfo {year} {2013})}\BibitemShut {NoStop}%
\bibitem [{Note1()}]{Note1}%
  \BibitemOpen
  \bibinfo {note} {Our use of $\chi _{\perp }$ and $\chi _{\parallel }$
  corresponds to generally accepted for other types of magnetic materials. For
  MnSi $\chi _{\parallel }$ corresponds to $\protect \bf {H}$ parallel to
  (111)-plane, containing magnetic moments, i.e., $\protect \bf {H} \perp
  \protect \bf {Q}$. In some papers, see, e.g., Ref.\protect \,\protect
  \rev@citealpnum {PhysRevB.73.174402}, $\chi _{\parallel }$ corresponds to
  $\protect \bf {H} \parallel \protect \bf {Q}$}\BibitemShut {NoStop}%
\bibitem [{\citenamefont {R\"o\ss{}ler}\ \emph {et~al.}(2006)\citenamefont
  {R\"o\ss{}ler}, \citenamefont {Bogdanov},\ and\ \citenamefont
  {Pfleiderer}}]{Rosler_2006}%
  \BibitemOpen
  \bibfield  {author} {\bibinfo {author} {\bibfnamefont {U.~K.}\ \bibnamefont
  {R\"o\ss{}ler}}, \bibinfo {author} {\bibfnamefont {A.~N.}\ \bibnamefont
  {Bogdanov}}, \ and\ \bibinfo {author} {\bibfnamefont {C.}~\bibnamefont
  {Pfleiderer}},\ }\href@noop {} {\bibfield  {journal} {\bibinfo  {journal}
  {Nature}\ }\textbf {\bibinfo {volume} {442}},\ \bibinfo {pages} {797}
  (\bibinfo {year} {2006})}\BibitemShut {NoStop}%
\bibitem [{\citenamefont {Stishov}\ \emph {et~al.}(2008)\citenamefont
  {Stishov}, \citenamefont {Petrova}, \citenamefont {Khasanov}, \citenamefont
  {Panova}, \citenamefont {Shikov}, \citenamefont {Lashley}, \citenamefont
  {Wu},\ and\ \citenamefont {Lograsso}}]{Stishov_2008}%
  \BibitemOpen
  \bibfield  {author} {\bibinfo {author} {\bibfnamefont {S.~M.}\ \bibnamefont
  {Stishov}}, \bibinfo {author} {\bibfnamefont {A.~E.}\ \bibnamefont
  {Petrova}}, \bibinfo {author} {\bibfnamefont {S.}~\bibnamefont {Khasanov}},
  \bibinfo {author} {\bibfnamefont {G.~K.}\ \bibnamefont {Panova}}, \bibinfo
  {author} {\bibfnamefont {A.~A.}\ \bibnamefont {Shikov}}, \bibinfo {author}
  {\bibfnamefont {J.~C.}\ \bibnamefont {Lashley}}, \bibinfo {author}
  {\bibfnamefont {D.}~\bibnamefont {Wu}}, \ and\ \bibinfo {author}
  {\bibfnamefont {T.~A.}\ \bibnamefont {Lograsso}},\ }\href@noop {} {\bibfield
  {journal} {\bibinfo  {journal} {J. Phys.: Condens. Matter}\ }\textbf
  {\bibinfo {volume} {20}},\ \bibinfo {pages} {235222} (\bibinfo {year}
  {2008})}\BibitemShut {NoStop}%
\bibitem [{\citenamefont {Matsunaga}\ \emph {et~al.}(1982)\citenamefont
  {Matsunaga}, \citenamefont {Ishikawa},\ and\ \citenamefont
  {Nakajima}}]{Matsunaga_1982}%
  \BibitemOpen
  \bibfield  {author} {\bibinfo {author} {\bibfnamefont {M.}~\bibnamefont
  {Matsunaga}}, \bibinfo {author} {\bibfnamefont {Y.}~\bibnamefont {Ishikawa}},
  \ and\ \bibinfo {author} {\bibfnamefont {T.}~\bibnamefont {Nakajima}},\
  }\href@noop {} {\bibfield  {journal} {\bibinfo  {journal} {J. Phys. Soc.
  Japan}\ }\textbf {\bibinfo {volume} {51}},\ \bibinfo {pages} {1153} (\bibinfo
  {year} {1982})}\BibitemShut {NoStop}%
\bibitem [{\citenamefont {Thessieu}\ \emph {et~al.}(1997)\citenamefont
  {Thessieu}, \citenamefont {Pfleiderer}, \citenamefont {Stepanov},\ and\
  \citenamefont {Flouquet}}]{Thessieu_1997}%
  \BibitemOpen
  \bibfield  {author} {\bibinfo {author} {\bibfnamefont {C.}~\bibnamefont
  {Thessieu}}, \bibinfo {author} {\bibfnamefont {C.}~\bibnamefont
  {Pfleiderer}}, \bibinfo {author} {\bibfnamefont {A.~N.}\ \bibnamefont
  {Stepanov}}, \ and\ \bibinfo {author} {\bibfnamefont {J.}~\bibnamefont
  {Flouquet}},\ }\href@noop {} {\bibfield  {journal} {\bibinfo  {journal} {J.
  Phys.: Condens. Matter}\ }\textbf {\bibinfo {volume} {9}},\ \bibinfo {pages}
  {6677} (\bibinfo {year} {1997})}\BibitemShut {NoStop}%
\bibitem [{\citenamefont {Grigoriev}\ \emph
  {et~al.}(2006{\natexlab{b}})\citenamefont {Grigoriev}, \citenamefont
  {Maleyev}, \citenamefont {Okorokov}, \citenamefont {Chetverikov},\ and\
  \citenamefont {Eckerlebe}}]{PhysRevB.73.224440}%
  \BibitemOpen
  \bibfield  {author} {\bibinfo {author} {\bibfnamefont {S.~V.}\ \bibnamefont
  {Grigoriev}}, \bibinfo {author} {\bibfnamefont {S.~V.}\ \bibnamefont
  {Maleyev}}, \bibinfo {author} {\bibfnamefont {A.~I.}\ \bibnamefont
  {Okorokov}}, \bibinfo {author} {\bibfnamefont {Y.~O.}\ \bibnamefont
  {Chetverikov}}, \ and\ \bibinfo {author} {\bibfnamefont {H.}~\bibnamefont
  {Eckerlebe}},\ }\href {\doibase 10.1103/PhysRevB.73.224440} {\bibfield
  {journal} {\bibinfo  {journal} {Phys. Rev. B}\ }\textbf {\bibinfo {volume}
  {73}},\ \bibinfo {pages} {224440} (\bibinfo {year}
  {2006}{\natexlab{b}})}\BibitemShut {NoStop}%
\bibitem [{\citenamefont {Bizette}\ and\ \citenamefont
  {Tsai}(1954)}]{Bizette_1954}%
  \BibitemOpen
  \bibfield  {author} {\bibinfo {author} {\bibfnamefont {H.}~\bibnamefont
  {Bizette}}\ and\ \bibinfo {author} {\bibfnamefont {B.}~\bibnamefont {Tsai}},\
  }\href@noop {} {\bibfield  {journal} {\bibinfo  {journal} {Compt. Rend.}\
  }\textbf {\bibinfo {volume} {238}},\ \bibinfo {pages} {1575} (\bibinfo {year}
  {1954})}\BibitemShut {NoStop}%
\bibitem [{\citenamefont {Jagadeesh}\ and\ \citenamefont
  {Seehra}(1981)}]{PhysRevB.23.1185}%
  \BibitemOpen
  \bibfield  {author} {\bibinfo {author} {\bibfnamefont {M.~S.}\ \bibnamefont
  {Jagadeesh}}\ and\ \bibinfo {author} {\bibfnamefont {M.~S.}\ \bibnamefont
  {Seehra}},\ }\href {\doibase 10.1103/PhysRevB.23.1185} {\bibfield  {journal}
  {\bibinfo  {journal} {Phys. Rev. B}\ }\textbf {\bibinfo {volume} {23}},\
  \bibinfo {pages} {1185} (\bibinfo {year} {1981})}\BibitemShut {NoStop}%
\bibitem [{\citenamefont {Cho}\ \emph {et~al.}(2005)\citenamefont {Cho},
  \citenamefont {Rhyee}, \citenamefont {Kim}, \citenamefont {Emilia},\ and\
  \citenamefont {Canfield}}]{Cho_2005}%
  \BibitemOpen
  \bibfield  {author} {\bibinfo {author} {\bibfnamefont {B.~K.}\ \bibnamefont
  {Cho}}, \bibinfo {author} {\bibfnamefont {J.-S.}\ \bibnamefont {Rhyee}},
  \bibinfo {author} {\bibfnamefont {J.~Y.}\ \bibnamefont {Kim}}, \bibinfo
  {author} {\bibfnamefont {M.}~\bibnamefont {Emilia}}, \ and\ \bibinfo {author}
  {\bibfnamefont {P.~C.}\ \bibnamefont {Canfield}},\ }\href@noop {} {\bibfield
  {journal} {\bibinfo  {journal} {J. Appl. Phys.}\ }\textbf {\bibinfo {volume}
  {97}},\ \bibinfo {pages} {10A923} (\bibinfo {year} {2005})}\BibitemShut
  {NoStop}%
\bibitem [{\citenamefont {Sato}\ \emph {et~al.}(2000)\citenamefont {Sato},
  \citenamefont {Enoki}, \citenamefont {Isobe},\ and\ \citenamefont
  {Ueda}}]{PhysRevB.61.3563}%
  \BibitemOpen
  \bibfield  {author} {\bibinfo {author} {\bibfnamefont {H.}~\bibnamefont
  {Sato}}, \bibinfo {author} {\bibfnamefont {T.}~\bibnamefont {Enoki}},
  \bibinfo {author} {\bibfnamefont {M.}~\bibnamefont {Isobe}}, \ and\ \bibinfo
  {author} {\bibfnamefont {Y.}~\bibnamefont {Ueda}},\ }\href {\doibase
  10.1103/PhysRevB.61.3563} {\bibfield  {journal} {\bibinfo  {journal} {Phys.
  Rev. B}\ }\textbf {\bibinfo {volume} {61}},\ \bibinfo {pages} {3563}
  (\bibinfo {year} {2000})}\BibitemShut {NoStop}%
\bibitem [{\citenamefont {Johnston}(2012)}]{PhysRevLett.109.077201}%
  \BibitemOpen
  \bibfield  {author} {\bibinfo {author} {\bibfnamefont {D.~C.}\ \bibnamefont
  {Johnston}},\ }\href {\doibase 10.1103/PhysRevLett.109.077201} {\bibfield
  {journal} {\bibinfo  {journal} {Phys. Rev. Lett.}\ }\textbf {\bibinfo
  {volume} {109}},\ \bibinfo {pages} {077201} (\bibinfo {year}
  {2012})}\BibitemShut {NoStop}%
\bibitem [{\citenamefont {Johnston}(2015)}]{arXiv:1407.6353}%
  \BibitemOpen
  \bibfield  {author} {\bibinfo {author} {\bibfnamefont {D.~C.}\ \bibnamefont
  {Johnston}},\ }\href {\doibase 10.1103/PhysRevB.91.064427} {\bibfield
  {journal} {\bibinfo  {journal} {Phys. Rev. B}\ }\textbf {\bibinfo {volume}
  {91}},\ \bibinfo {pages} {064427} (\bibinfo {year} {2015})}\BibitemShut
  {NoStop}%
\end{thebibliography}%

\end{document}